\documentclass[hyper]{andp2012}% no class options needed by now

\usepackage[english]{babel}
\usepackage{bbm}

\newcommand{\av}[1]{\langle #1\rangle}
\title{Dephasing enhanced spin transport in the ergodic phase of a many-body localizable system}
\author[M. \v Znidari\v c]{Marko \v Znidari\v c\inst{1,}\footnote{Corresponding author\quad E-mail:~\textsf{marko.znidaric@fmf.uni-lj.si}}}
\author[J.\,J. Mendoza-Arenas]{Juan Jose Mendoza-Arenas\inst{2,3}}
\author[S.\,R. Clark]{Stephen R. Clark\inst{4,5}}
\author[J. Goold]{John Goold\inst{6}}
\address[1]{Physics Department, Faculty of Mathematics and Physics, University of Ljubljana, 1000 Ljubljana, Slovenia}
\address[2]{Departamento de F\'isica, Universidad de los Andes, A.A. 4976, Bogot\'a D. C., Colombia}
\address[3]{Clarendon Laboratory, University of Oxford, Parks Road, Oxford OX1 3PU, United Kingdom}
\address[4]{Department of Physics, University of Bath, Claverton Down, Bath BA2 7AY, United Kingdom}
\address[5]{Max Planck Institute for the Structure and Dynamics of Matter, University of Hamburg CFEL, Hamburg, Germany}
\address[6]{The Abdus Salam ICTP, Strada Costiera 11, 34151, Trieste, Italy}

\shortauthors{M. \v Znidari\v c et al.}
\begin{abstract}
We study high temperature spin transport in a disordered Heisenberg chain in the ergodic regime when bulk dephasing is present. We find that while dephasing always renders the transport diffusive, there is nonetheless a remnant of the diffusive to sub-diffusive transition found in a system without dephasing manifested in the behaviour of the diffusion constant with the dephasing strength. By studying finite-size effects we show numerically and theoretically that this feature is caused by the competition between large crossover length scales associated to disorder and dephasing that control the dynamics observed in the thermodynamic limit. We demonstrate that this competition may lead to a dephasing enhanced transport in this model. 
\end{abstract}
\shortabstract

\begin{document}
\maketitle

\section{Introduction}

The simple model of a particle hopping on a lattice in the presence of disorder represents an iconic system in condensed matter physics. The consideration of this model lead to the realisation by Anderson that a static disordered potential can lead to a complete absence of diffusion in an isolated quantum system due to the localisation of electronic states, resulting in a non-ergodic and insulating phase. This phenomenon is known as \emph{Anderson localization} \cite{Anderson:1958}. Although discovered more than half a century ago, this single particle problem is still inspiring exciting research avenues. Surprisingly, the role of interactions has come under intense investigation only recently as it was generally believed that interactions destabilised localisation due to resonance effects. However in a seminal contribution in 2006, Basko, Aleiner and Altshuler have shown using a perturbative argument that Anderson localisation is stable in the presence of interactions \cite{Basko:2006} leading to a new type of phase known as many-body localization (MBL), which is currently under intense theoretical \cite{Nandkishore:2015,Altman:2015} and experimental investigation \cite{Schreiber:2015, Smith:2016, Bordia:2016, Luschen:2016}.

Following the initial discovery, early numerical studies on one dimensional spin systems highlighted not only the interesting properties of the MBL phase itself but also the highly non trivial nature of the phase transition present at infinite temperature \cite{Oganesyan:2007,Pal:2010}. Over the past decade studies have uncovered a rich variety of phenomenology in particular focusing on the MBL phase \cite{Nandkishore:2015,Altman:2015}. Logarithmic growth of entanglement \cite{Znidaric:2008}, the emergence of extensive set of quasi conserved quantities \cite{Serbyn:2013b,Ros:2015,Imbrie:2016}, the existence of a finite energy density mobility edge \cite{Luitz:2015}, protection of interesting states and phases~\cite{Huse:2013,Chandran:2014,Vasseur:2015}, and changes in the correlations of the system \cite{Bauer:2013,Friesdorf:2015,Goold:2015,Lemini:2016,Campbell:2016} are just some of the wide range of interesting phenomenology that has been discovered. The picture, at least deep within the MBL phase, provided it exists (as in one-dimensional systems with short-range interactions), is by now well understood.

However, far less is known about the system within the ergodic phase and in the transition region. In particular, due to the small system sizes reached in most numerical studies (of order of 20 spins), the transport properties of the 
ergodic phase are still under active investigation and have led to conflicting results regarding the existence and size of a sub-diffusive phase \cite{Lev:2015,Lucioni:2011,Gopalakrishnan:2015,Agarwal2015,Dalmonte:2015,Gopalakrishnan:2016,Luitz:2016,Varma:2015}. To correctly predict physical properties in the thermodynamic limit it is absolutely crucial to study large systems, for which a boundary driven Lindblad setup seems to be the best framework, enabling one to reach systems of several hundred sites~\cite{Znidaric:2016}. Using this technique, in a system without dephasing a diffusive phase has been identified at low disorder followed by a sharp crossover to a sub-diffusive phase. 

Another area of active investigation corresponds to the dynamical and steady state features which emerge when a disordered and interacting system is coupled to an external bath \cite{Nandkishore:2014,Johri:2015,Medvedyeva:2016,Levi:2016,Fischer:2016,Nandkishore:2016,Everest:2016}. These works have primarily considered systems which are MBL in the absence of the bath and have focused on the degradation of the phase by bath induced dephasing. In this work we are primarily interested on the effect of bath dephasing on the transition between diffusive and sub-diffusive regions of the ergodic phase identified in~\cite{Znidaric:2016}. The setup is described by Lindblad terms acting on both the bulk to describe dephasing and at the boundaries to model transport. Markovian dephasing generically renders all transport diffusive in the thermodynamic limit, even in the presence of disorder and/or interactions, see e.g.~\cite{Znidaric:2010,Mendoza:2013a,Mendoza:2013b} (for noninteracting model~\footnote{We note that even a correlated noise (Markovian dephasing being a limit of an uncorrelated noise) renders transport diffusive in a noninteracting model~\cite{Knap:2016}.} exact solutions are possible~\cite{Znidaric:2010b,Horvat:2013,Stegmann:2014}) we are here interested in how this diffusive behavior comes about. In particular, we find remnants of the diffusive to sub-diffusive transition in the behaviour of the diffusion coefficient as a function of the dephasing parameter. We identify two length scales in this model, one corresponding to dephasing and the other to disorder. We show that the competition between these two scales can lead to a dephasing enhanced spin transport when disorder is strong enough to cause sub-diffusive transport in the absence of dephasing.

\section{System}
\begin{figure} 
  \includegraphics*[width=\columnwidth]{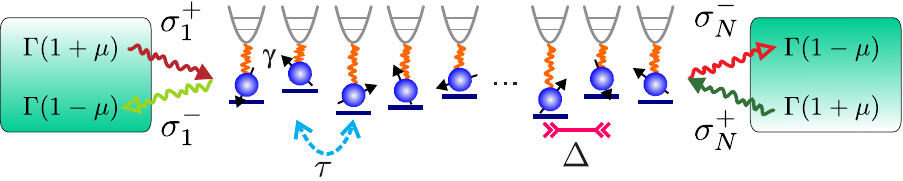} 
  \caption{Scheme of a boundary-driven disordered $XXZ$ spin chain. Spin excitations can hop between neighbouring lattice sites with amplitude $\tau$, and experience Ising-type interactions of strength $\tau\Delta$; disorder is depicted as random on-site energies. Boundary reservoirs inject and eject excitations at each end of the chain, with $\mu$ determining the imbalance between both processes. Each site experiences dephasing with rate $\gamma$, arising e.g. from coupling to local vibrational degrees of freedom.}
  \label{fig1_setup}
\end{figure}

We consider here an open-boundary spin-chain composed of $L$ spins governed by the anisotropic Heisenberg model with disorder in the longitudinal fields. This is given as $\hat{H} = \tau \sum_{l=1}^{L-1} \hat{h}_{l,l+1}$, with (we set the energy scale $\tau = 1$)
\begin{equation} 
\hat{h}_{l,l+1} = \hat{\sigma}^x_l\hat{\sigma}^x_{l+1} + \hat{\sigma}^y_l\hat{\sigma}^y_{l+1} + \Delta\hat{\sigma}^z_l\hat{\sigma}^z_{l+1} + \frac{h_l}{2}\hat{\sigma}^z_l + \frac{h_{l+1}}{2}\hat{\sigma}^z_{l+1},
\end{equation}
where $\hat{\sigma}^\alpha_l$ are the $(\alpha=x,y,z)$ spin-$1/2$ Pauli operators for the $l$th spin, $\Delta$ is the anisotropy, and $h_l \in [-h,h]$ is a uniformly random disorder of strength $h$. To study the nature of transport in this system we couple its ends to independent ``magnetic'' reservoirs, as depicted in Figure~\ref{fig1_setup}, that induce a spin-current carrying non-equilibrium steady state (NESS). Specifically, we model the dynamics of the system via a Lindblad master equation for its density matrix as \cite{Prosen:2009,Znidaric:2010,Znidaric:2010b,Prosen:2011,Horvat:2013,Mendoza:2013a,Mendoza:2013b,Karevski:2013,Landi:2015} (setting $\hbar=1$)
\begin{eqnarray}
\frac{\rm d}{{\rm d}t}\hat{\rho}(t) &=& {\rm i}[\hat{\rho}(t),\hat{H}] +  \mathcal{D}(\hat{L}^+_1)\{\hat{\rho}(t)\}  + \mathcal{D}(\hat{L}^-_1)\{\hat{\rho}(t)\} \label{master_eq} \\
&&  + \mathcal{D}(\hat{L}^+_L)\{\hat{\rho}(t)\} + \mathcal{D}(\hat{L}^-_L)\{\hat{\rho}(t)\} + \sum_{l=1}^{L} \mathcal{D}(\hat{L}^z_l)\{\hat{\rho}(t)\}, \nonumber
\end{eqnarray}
where $\mathcal{D}(\hat{L})\{\hat{\rho}\} = [\hat{L}\hat{\rho},\hat{L}^\dagger] + [\hat{L}, \hat{\rho}\hat{L}^\dagger]$ is the dissipation super-operator defined in terms of Lindblad jump operators $\hat{L}$. The boundary driving is described by Lindblad operators $L^+_1 = \sqrt{\Gamma(1+\mu)}~\hat{\sigma}^+_1$, $L^-_1 = \sqrt{\Gamma(1-\mu)}~\hat{\sigma}^-_1$ for the left end, and correspondingly $L^+_L = \sqrt{\Gamma(1-\mu)}~\hat{\sigma}^+_L$, $L^-_L = \sqrt{\Gamma(1+\mu)}~\hat{\sigma}^-_L$ at the right end, with $\hat{\sigma}_l^\pm = (\hat{\sigma}^x_l \pm \hat{\sigma}^y_l)/2$. Here $\mu$ controls the spin imbalance between the two baths and provided $\mu \neq 0$ a NESS with a spin current is induced. The last terms in Eq.~\ref{master_eq} describe bulk dephasing of each spin at a rate $\gamma$ given by Lindblad operators $\hat{L}^z_l = \sqrt{\gamma/2} ~\hat{\sigma}^z_l$.

This system has a unique NESS $\hat{\rho}_\infty$ and so any initial state $\hat{\rho}(0)$ eventually converges under the time-evolution described by Eq.~\ref{master_eq} as $\lim_{t\rightarrow\infty} \hat{\rho}(t) = \hat{\rho}_\infty$. Exploiting this we obtain $\hat{\rho}_\infty$ by simulating the real time evolution for times sufficiently long that this convergence is reached. This is performed with the time-dependent density matrix renormalisation group (t-DMRG) algorithm \cite{Zwolak:2004,Cirac:2004,Daley:2004,Alassam2016,Schollwock:2011} as we describe in the next section. The main observable of interest is the spin-current between spins $l-1$ and $l$ determined by the continuity equation ${\rm d} \hat{\sigma}^z_l/{\rm d} t = {\rm i}[\hat{\sigma}^z_l,\hat{H}]  = \hat{j}_{l} - \hat{j}_{l - 1}$ as $\hat{j}_l = 2(\hat{\sigma}^x_l\hat{\sigma}^y_{l+1} - \hat{\sigma}^y_l\hat{\sigma}^x_{l+1})$. Using the NESS we then compute $\av{\hat{j}_l} = {\rm tr}(\hat{\rho}_\infty \hat{j}_l)$, which due to stationarity is independent of $l$ and will be denoted as $j$. Our results are then ensemble averaged over disorder realisations $h_l$ with a number $M$ of samples chosen to ensure that the statistical uncertainty in $j$ is less than 2\%. Simulations also reach system sizes up to $L=400$, which is essential to accurately uncover the transport scaling of $j$ with $L$ in the thermodynamic limit. 

The focus of our work is fixed on the isotropic point $\Delta = 1$, which despite being studied intensively still displays behaviour that is not fully understood. We also limit ourselves to disorder strengths in the ergodic phase below the MBL transition point at $h_{c3} \approx 7.4$~\cite{Pal:2010,Luitz:2015}. To analyse the transport we work with strong coupling $\Gamma = 1$ in the linear response regime $\mu = 0.001$ where $\hat{\rho}_\infty$ is close to the infinite temperature NESS with no driving $\hat{\rho}_\infty(\mu=0) \sim \mathbbm{1}$,  where $\mathbbm{1}$ is the identity matrix. The form of driving imposed in Eq.~\eqref{master_eq} allows us to exclusively focus on spin transport $j$, since the disorder-averaged energy current is zero \cite{Znidaric:2016}.  

Importantly, recent seminal advances in the manipulation and measurement of quantum simulators with cold atomic gases would allow the study of non-equilibrium systems similar to that described in the present work in the laboratory. In particular, the combination of simulation schemes of particle transport through low-dimensional channels connecting unequal fermionic reservoirs \cite{Brantut:2012,Stadler:2012,Krinner:2013,Krinner:2015} and the possibility of implementing spin Hamiltonians in fermionic optical lattices \cite{Greif:2013,Greif:2015} would provide an ideal experimental setup to test the predictions of our work.

\section{Method}
\begin{figure}
  \includegraphics*[width=7cm]{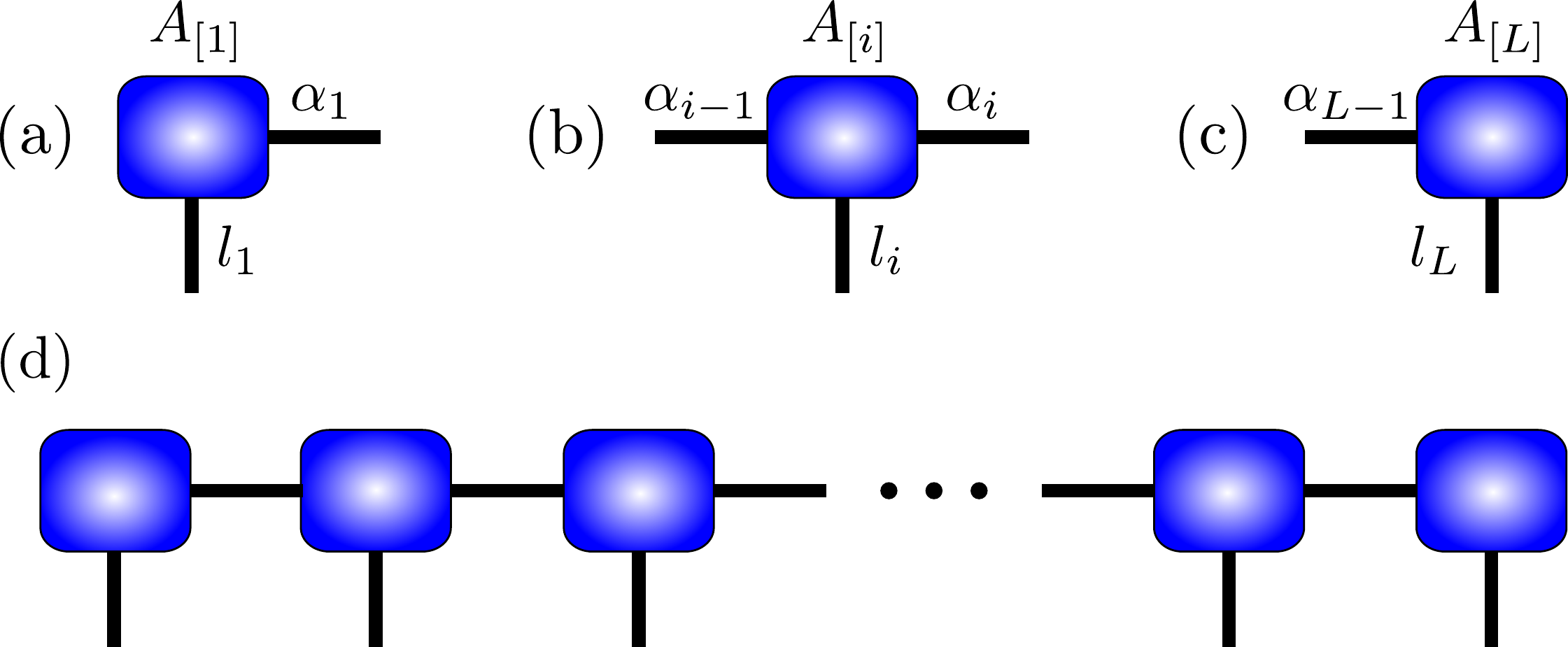} 
  \caption{Graphical representation of MPO formalism. The building blocks of an MPO are single-site tensors, shown for the left boundary, (b) the bulk, and (c) the right boundary. The vertical lines correspond to the physical indices. The horizontal lines represent the rows and columns of each tensor, of dimension $\chi$ and indices $\alpha_l$. (d) Graphical representation of the MPO structure of $\hat{\rho}$.}
  \label{mpo_graph}
\end{figure}

In this Section we review the basic details of the t-DMRG method used to obtain the NESS of the open spin chain. For this we describe the density matrix of the lattice $\hat{\rho}(t)$ by a Matrix Product Operator (MPO) structure~\cite{Zwolak:2004,Prosen:2009,Schollwock:2011} 
\begin{equation} \label{MPO}
\hat{\rho}=\sum_{\ell_1,\ell_2,\ldots,\ell_L=1}^{d}A_{[1]}^{\ell_1}A_{[2]}^{\ell_2}\cdots A_{[L]}^{\ell_L}\left(\hat{\sigma}_{1}^{\ell_1}\otimes\hat{\sigma}_{2}^{\ell_2}\otimes\cdots\otimes\hat{\sigma}_{L}^{\ell_L}\right),
\end{equation}
with $\hat{\sigma}_l^{1}=\frac{1}{2}\mathbbm{I}$ (normalised $2\times2$ identity) and $\hat{\sigma}_\ell^{2,3,4}=\hat{\sigma}_{l}^{x,y,z}$, and where the information of site $l$ is contained in the $d-$rank tensor $A_{[l]}$, formed by $d=4$ matrices $A_{[l]}^{\ell_l}$ of dimension $\chi\times\chi$. This structure has a nice graphical representation~\cite{Schollwock:2011}, where each tensor is depicted as a block with several legs (see Figure~\ref{mpo_graph}). The entire MPO is built considering that tensors are contracted when connected by a common leg. Thus the operation specified in Eq.~\eqref{MPO} is represented by Figure~\ref{mpo_graph}(d). This graphical notation allows us to describe the operations required to perform time evolution in a simple way.

\begin{figure}[b]
  \includegraphics*[width=7cm]{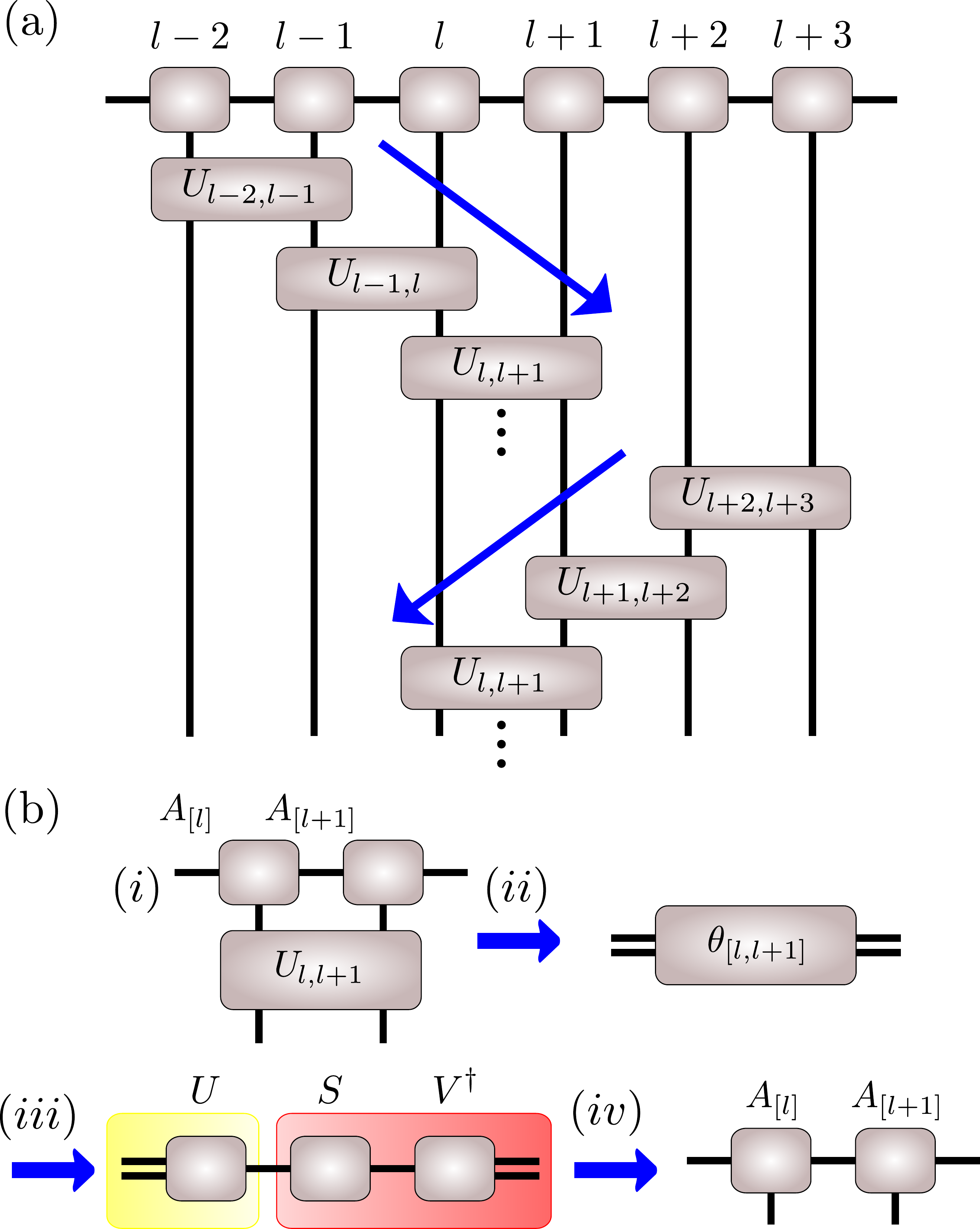} 
  \caption{(a) t-DMRG sweep process. At each time step, the evolution operator is approximated by a left-to-right sweep of ordered contractions of the MPO with local two-site gates $U_{l,l+1}$, followed by a right to-left sweep. (b) Application of a two-site gate to an MPO. (i) The tensors of the two sites are contracted with the gate. (ii) The MPO structure is lost by this contraction. (iii) Perform a SVD. (iv) Recover the MPO structure by reshaping the matrices resulting from the SVD. $U$ forms the matrices of site $l$, and $SV^{\dagger}$ of site $l+1$.}
  \label{tebd}
  
\end{figure}
For this, we write the equation of motion of $\hat{\rho}(t)$ as
\begin{equation}
\frac{\rm d}{{\rm d}t}\hat{\rho}(t)=\mathcal{L}\{\hat{\rho}(t)\}=\sum_{l=1}^{L-1}\mathcal{L}_{l,l+1}\{\hat{\rho}(t)\},
\end{equation}
where $\mathcal{L}$ represents the total dissipator of the system and $\mathcal{L}_{l,l+1}$ the dissipator acting on sites $l,l+1$. Thus the state at time $t$ is obtained from the initial condition as $\hat{\rho}(t)=\exp(\mathcal{L}t)\{\hat{\rho}(0)\}$. This state is reached by dividing the total time domain into $T$ steps of size $\delta t$, so that
\begin{equation}
\exp(\mathcal{L}t)=\prod_{m=1}^T\exp(\mathcal{L}\delta t)=[\exp(\mathcal{L}\delta t)]^{t/\delta t},
\end{equation}
with $\delta t$ small enough so the fastest energy scale of the system is captured. Furthermore, the evolution operator at each time step is approximated by a Suzuki-Trotter decomposition~\cite{suzuki1990fractal}. For instance, up to second order
\begin{align} \label{suzuki}
\begin{split}
\exp(\mathcal{L}\delta t)&=\Biggl(\prod_{l=1}^{L-1}\exp(\mathcal{L}_{l,l+1}\delta t/2)\Biggr)\\
&\times\Biggl(\prod_{l=L-1}^{1}\exp(\mathcal{L}_{l,l+1}\delta t/2)\Biggr)+O(\delta t^3)
\end{split}
\end{align}
In this form, the global evolution operator is reduced to two sweeps, one from left to right and other from right to left, each consisting of two-site gates $U_{l,l+1}=\exp(\mathcal{L}_{l,l+1}\delta t/2)$ applied in ordered sequence, as illustrated in Figure~\ref{tebd}(a).

The application of a two-site gate to the MPO is depicted in Figure~\ref{tebd}(b). After the initial contraction, the MPO structure is lost. However using a singular value decomposition (SVD) of the resulting tensor $\theta$, the MPO is recovered. The SVDs also help controlling the size of the resulting matrices $A_{[l]}^{\ell_l}$, which in the absence of any approximation would typically grow in time. This is achieved by setting their dimension to a fixed $\chi$, with a total error given by the sum of the discarded elements of the matrix $S$ of each SVD (see figure~\ref{tebd}(b)). In our case, moderate values of $\chi$ are needed to obtain accurate results, using up to $\chi=150$ for the most expensive simulations.

\section{Scaling}
In the ergodic phase ($h < h_{c3}$) there is non-zero spin transport. It is well known that for microscopic models of transport the mean square displacement $\Delta x$ of a particle (or spin inhomogeneity) grows asymptotically in time as 
\begin{equation}
\av{\Delta x^2} \sim t^{2\alpha},
\end{equation} 
where $ 0 < \alpha \leq 1$. Another characteristic transport property is the scaling of the current in the steady state at fixed driving and which is expected to have the form
\begin{equation}
j \sim 1/L^\nu,
\end{equation} 
where $\nu \geq 0$. From a simple single-parameter scaling analysis, arguing that the transition time across the system for a single excitation is $\sim L^{1/\alpha}$, so that the current at fixed density scales as $j \sim L/L^{1/\alpha}$, the two exponents are related as
\begin{equation}
\alpha = 1/(\nu + 1),
\end{equation}
and together classify the possible behaviour of the system. Normal {\em diffusive} transport corresponds to $\nu = 1$ where $\av{\Delta x^2}$ grows linearly in time and $j \sim 1/L$ obeying the phenomenological transport law $j = -D\nabla\av{\hat{\sigma}^z_\ell}$. For our driving the average gradient for large systems is $\nabla\av{\hat{\sigma}^z_\ell}=2\mu/L$. Other values of $\nu$ describe anomalous transport in which this equation is obeyed so long as $D$ is allowed to be length dependent as $D \sim L^{1-\nu}$. When $\nu > 1$ we have {\em sub-diffusive} transport where the variance $\av{\Delta x^2}$ grows slower than $t$, $j$ decays faster with $L$ than diffusion, and $D$ vanishes in the thermodynamic limit. The extreme limit of this is $\nu \rightarrow \infty$ where $\av{\Delta x^2}$ does not grow at all, symptomatic of localisation. At such point the power-law scaling of the current breaks down and we instead have the emergence of insulating behaviour $j \sim \exp(-L/L_0)$ , e.g. as encountered when entering the MBL phase ($h > h_{c3}$). When $\nu < 1$ we have {\em super-diffusive} transport where $\av{\Delta x^2}$ grows faster than $t$, the current decays slower with $L$ than diffusion, and $D$ diverges in the thermodynamic limit. In the extreme case $\nu = 0$ we have {\em ballistic} transport where $\av{\Delta x^2} \sim t^2$ consistent with a freely propagating particle, $j$ is independent of $L$ and $D \sim L$.  

In the absence of dephasing it has been found~\cite{Znidaric:2016} that at $\Delta=1$ the asymptotic transport behaviour induced by disorder is diffusive for $h < h_{c2} \approx 1.1$, and sub-diffusive for $h > h_{c2}$, while it is super-diffusive for $h_{\rm c1}=0$~\cite{Znidaric:2011b}. In the diffusive case and for small disorder there emerges a length scale $L_*$,
\begin{equation}
L_* \sim h^{-2/(\nu+1)},
\label{eq:L*}
\end{equation}
above which the diffusive behaviour sets in (with $2/(\nu+1) = 1.33$ at $\Delta=1$ and $h \to 0$). The introduction of any non-zero dephasing is known to render transport diffusive for any $h$ on asymptotically long length scales. Our interest here is on the analogous crossover length scale $L_\gamma$ for $\gamma \ll 1$ where this dephasing-induced diffusive behaviour sets in. In other words, we want to understand how a small dephasing affects transport properties of a Hamiltonian, i.e., noiseless system. We estimate $L_\gamma$ by taking the time $\tau \sim 1/\gamma$ between the dephasing scattering events and computing the spatial spread $\sqrt{\av{\Delta x^2}}$ obtained within this time for evolution without dephasing. This gives a length scale
\begin{equation}
L_\gamma \sim \gamma^{-1/(\nu+1)},
\label{eq:Lgamma}
\end{equation}
beyond which dephasing will be important. Transport in a disordered and dephased system will depend delicately on $h$ and $\gamma$. Which contribution controls the behaviour is a question of competing length scales $L_*$ and $L_\gamma$, with the dominant one being the smallest. We now proceed to analyse this behaviour, first for a clean system to demonstrate the predicted behaviour of $L_\gamma$, and then for a disordered system for strengths $h$ above as well as below $h_{c2}$.

\section{Clean system} \label{clean}
\begin{figure} 
  \includegraphics*[width=0.95\columnwidth]{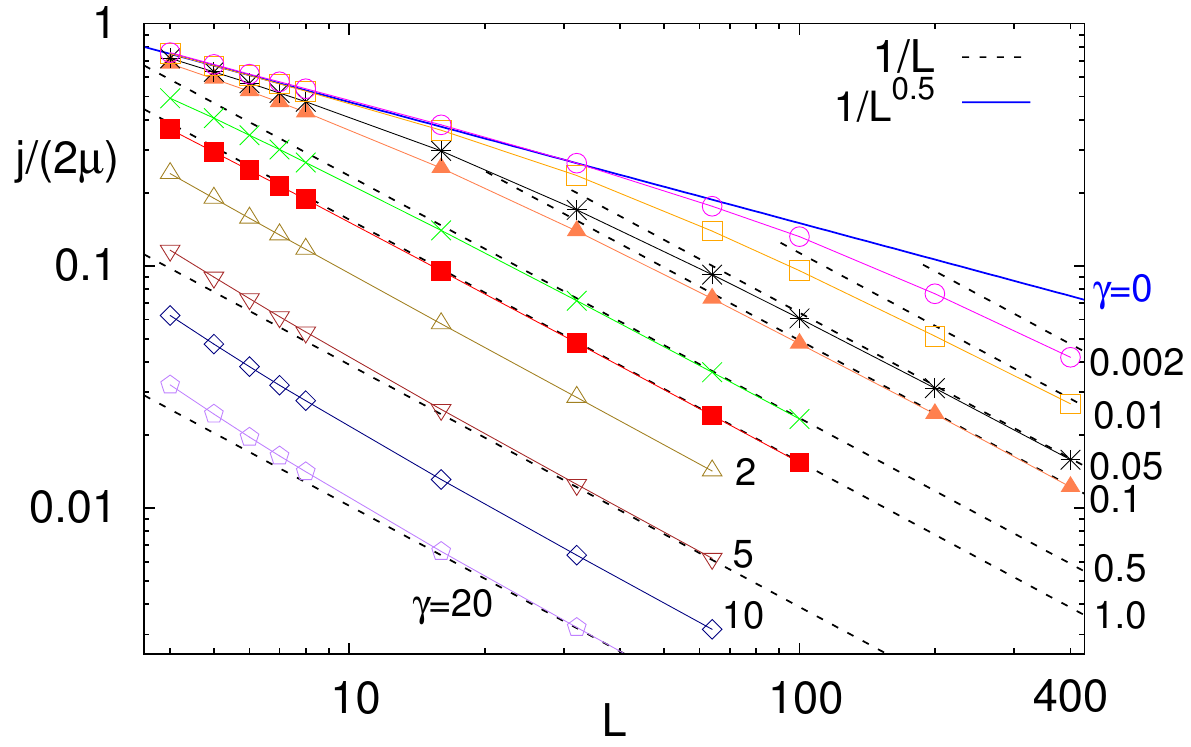} 
  \caption{The spin current $j$ scaling with $L$ for the clean, $h=0$, isotropic, $\Delta = 1$, Heisenberg chain for a variety of $\gamma$'s. For large $L> L_\gamma$ diffusion always sets in.}
  \label{fig1}
\end{figure}

We first consider transport in the clean system $h = 0$ (always with $\Delta=1$). This is known to be super-diffusive with $\nu = 1/2$~\cite{Znidaric:2011b} so we expect that $L_\gamma \sim \gamma^{-2/3}$. In Fig.~\ref{fig1} we show how the current $j$ scales with $L$ for a variety of $\gamma$'s. The behaviour clearly shows distinct short and long $L$ tendencies confirming the existence of a length scale $L_\gamma$ separating them. For small $\gamma$ and $L \ll L_\gamma$ the current scaling mimics the clean system with $j \sim 1/\sqrt{L}$, while for sufficiently large $L > L_\gamma$ the onset of $j \sim 1/L$ scaling is seen, confirming that diffusive transport is found. It is also apparent that for small dephasing rates $\gamma \leq 0.01$ even $L=400$ spins is only just sufficient to resolve this asymptotic scaling.

\begin{figure} 
  \includegraphics*[width=0.95\columnwidth]{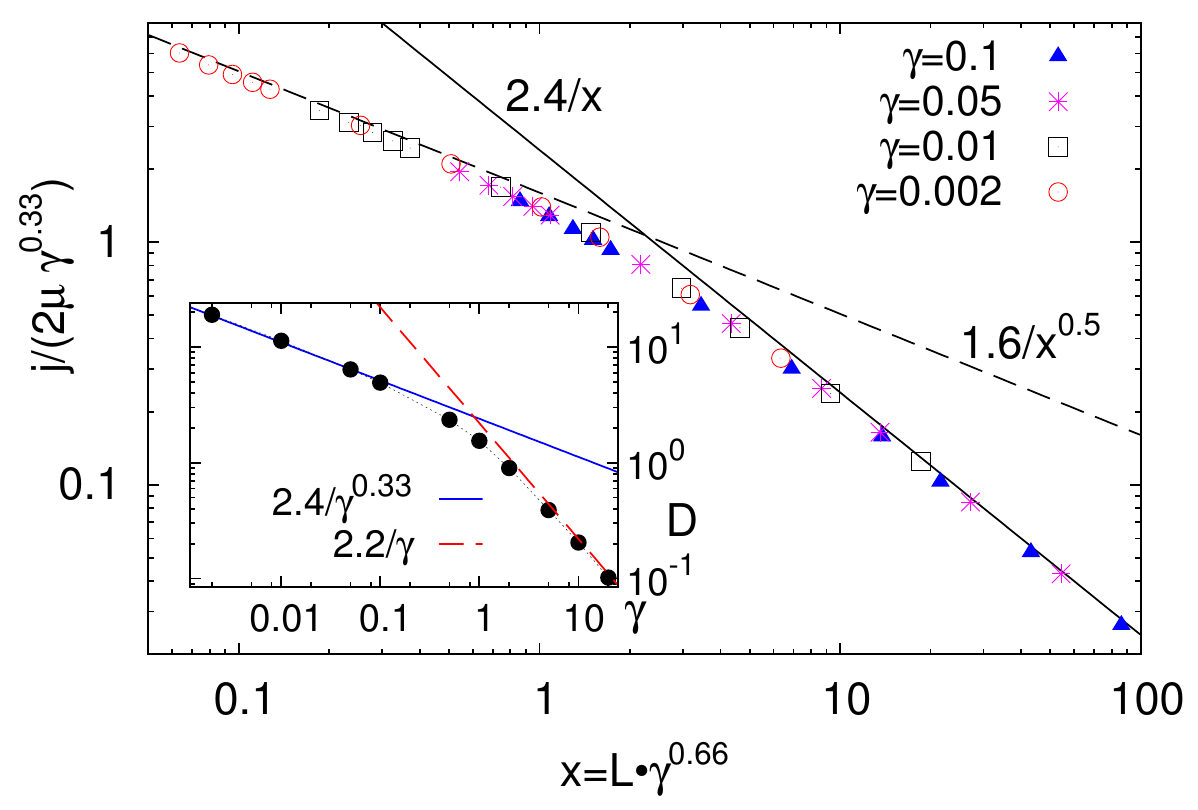} 
  \caption{The rescaled spin current $\tilde{j}$ against the rescaled length $x$ for the clean $h=0$ isotropic $\Delta = 1$ Heisenberg chain for a variety of small $\gamma$'s. The distinct $x$ dependences for the small and large $x$ regimes are highlighted by the fit lines for $1/x$ and $1/\sqrt{x}$. The inset shows $D$ as a function of $\gamma$ extracted from the largest system sizes where $L > L_\gamma$. The large and small regimes for $\gamma$ are highlighted by the fit lines for $1/\gamma^{1/3}$ from Eq.~\eqref{eq:smallg} (using $\nu=1/2$ valid at $\Delta=1$) and $1/\gamma$ expected for dephasing alone.}
  \label{fig2}
\end{figure}

Building on this observation we now consider more generally a clean system with transport properties characterised by an exponent $\nu$. The diffusion constant $D$ will be a function of both $L$ and $\gamma$. We model the presence of a crossover length scale $L_\gamma$ by taking $D$ to have a continuous piecewise dependence on $L$. For small $\gamma$, when $L_\gamma \gg 1$, the diffusion constant will take the form
\begin{equation}
D(\gamma,L) \sim
\left\{
\begin{array}{lcc}
L^{1-\nu},  && L < L_\gamma  \\
D_{\rm dph}(\gamma),  && L \geq L_\gamma
\end{array}
\right. . \label{diffusion_scaling}
\end{equation}
For $L <L_\gamma$ the anomalous transport of the clean system is dominant and $D$ is independent of $\gamma$. For $L \geq L_\gamma$ dephasing-induced diffusive transport sets in forcing $D = D_{\rm dph}(\gamma)$ independent of $L$. Continuity at $L_\gamma$ then implies that for small $\gamma$ 
\begin{equation}
D_{\rm dph}(\gamma) \sim L_\gamma^{1-\nu} \sim \gamma^{(\nu-1)/(\nu +1)},\qquad \gamma \ll 1.
\label{eq:smallg}
\end{equation}
The current then behaves as $j(\gamma,L) \sim D(\gamma,L)/L$. By rescaling the length as $x = L/L_\gamma$ and the current as $\tilde{j} = L_\gamma^\nu j = \gamma^{-\nu/(1+\nu)}j$ this gives a universal form
\begin{equation}
\tilde{j} \sim
\left\{
\begin{array}{lcc}
x^{-\nu},  && x \lessapprox 1  \\
x^{-1},  && x \gtrapprox 1
\end{array}
\right. . \label{current_scaling}
\end{equation}
In Fig.~\ref{fig2} we show the dependence of $\tilde{j}$ on $x$ for a variety of small $\gamma$'s. This plot demonstrates first the nice collapse of the data, and second the presence of the two regimes of scalings as given in Eq.~\eqref{current_scaling}. In the inset of Fig.~\ref{fig2} we show the extracted $D$ from the behaviour at lengths $L\sim400$ which for the $\gamma's$ examined is just long enough to see the diffusive regime. For small $\gamma$'s this plot confirms the $\gamma^{-1/3}$ scaling predicted by Eq.~\eqref{eq:smallg}. For large $\gamma$ the clean Hamiltonian influences only the very shortest length scales and transport is dephasing dominated with an expected dependence 
\begin{equation}
D_{\rm dph}(\gamma) \sim 1/\gamma,\qquad \gamma \gg 1,
\label{eq:largeg}
\end{equation}
coming from a perturbative expansion in $\frac{1}{\gamma}$.

\section{Disordered system}
Having established how $L_\gamma$ controls the behaviour of a clean dephased system we now proceed to examine its interplay with $L_*$ in a disordered system. A particularly interesting regime will be when both $L_\gamma$ given by Eq.~\eqref{eq:Lgamma} and $L_*$ given by Eq.\eqref{eq:L*} are large and one gets a competition between the two lengthscales. To study this we separately consider two cases depending on whether $h$ is smaller or larger than $h_{\rm c2}\approx 1.1$.

\subsection{Case (i) $h < h_{c2}$}
\begin{figure}[t!]
  \includegraphics*[width=0.96\columnwidth]{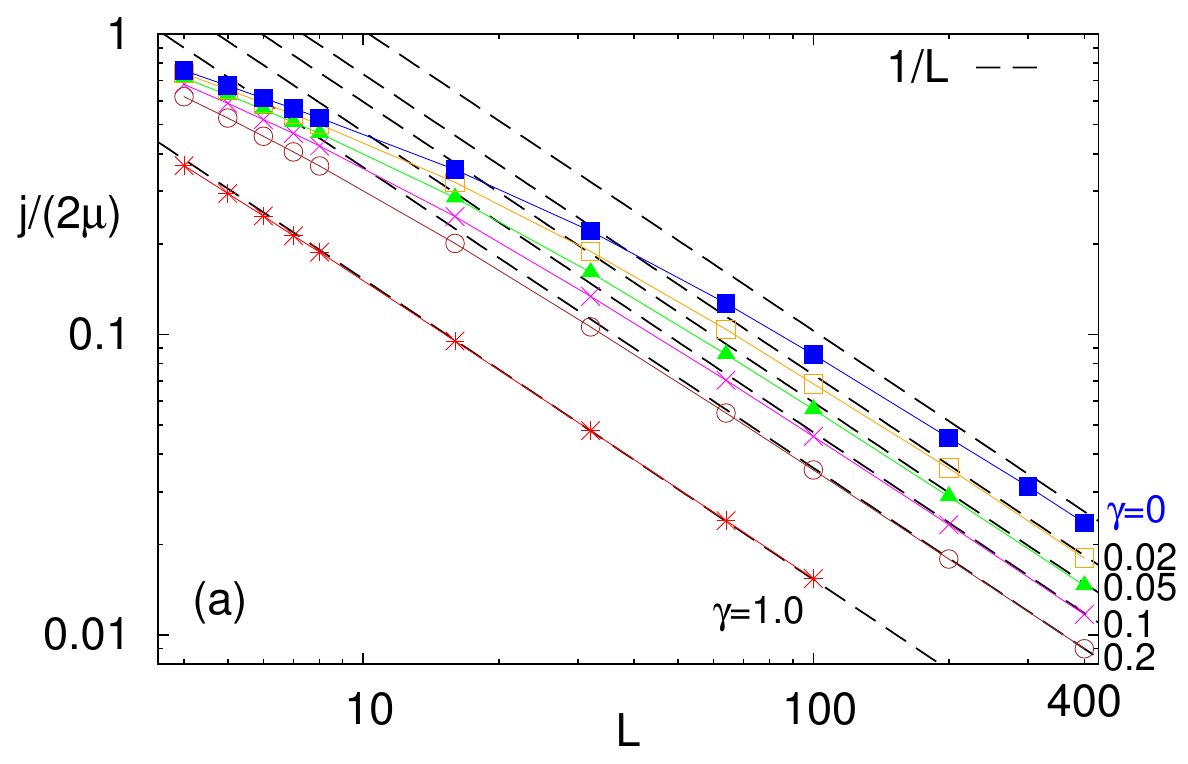} 
 \includegraphics*[width=0.9\columnwidth]{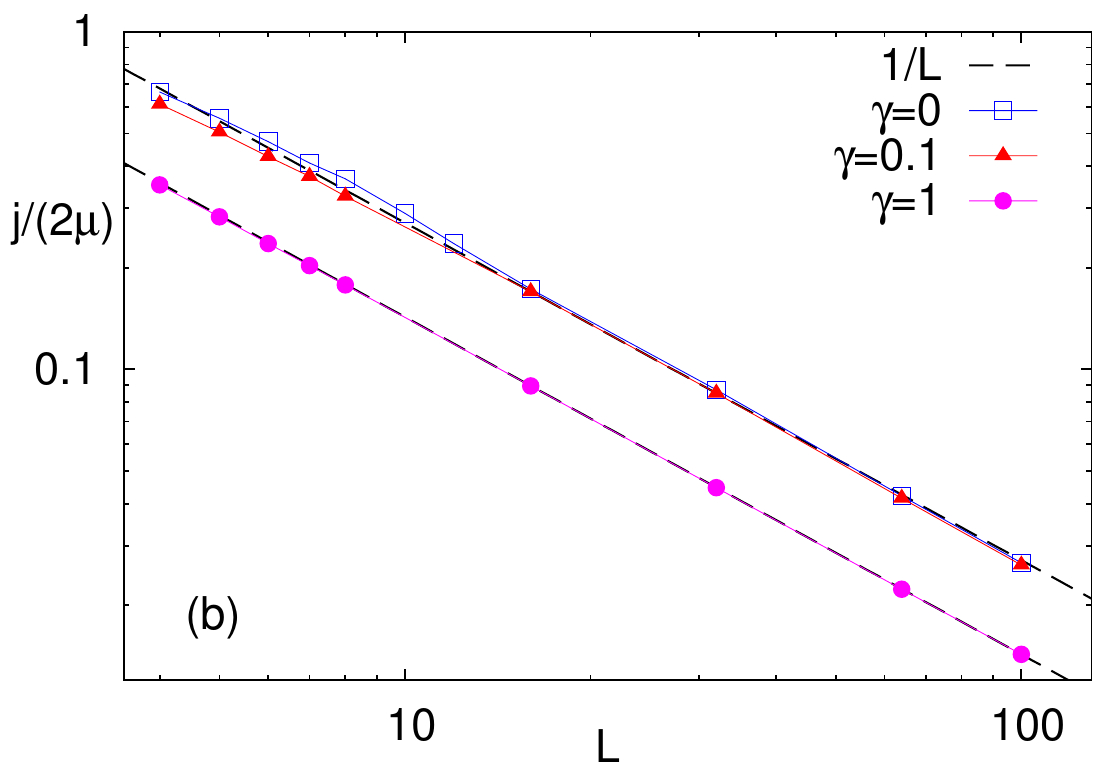} 
  \caption{For case $h<h_{\rm c}$ the current $j$ is shown as a function of $L$ for a variety of $\gamma$'s. (a) $h=0.2$ where $L_* > L_\gamma$ and one has dephasing-dominated diffusion. The crossover lengthscale $L_\gamma$ increases for decreasing $\gamma$. (b) $h=1.0$, where $L_*$ is small and, except for very large dephasing, diffusion is disorder-dominated, e.g., $\gamma=0.1$ is almost the same as $\gamma=0$.}
  \label{fig3}
\end{figure}

In Fig.~\ref{fig3} the current $j$ is shown as a function of $L$ for a variety of $\gamma$'s for (a) $h=0.2$ and (b) $h=1.0$, therefore fixing $L_*$ for each plot. For large $\gamma$ the $L_\gamma$ is the smallest length scale, i.e., disorder is just an irrelevant perturbation, and consequently $D$ behaves as in the clean system given by Eq.~\eqref{eq:largeg}, $D(\gamma) \sim 1/\gamma$, because diffusion is dephasing-dominated. As one decreases $\gamma$ the diffusion constant first increases, however, this increase does not diverge as $\gamma \rightarrow 0$, as it would for $h=0$. When $\gamma \lesssim h^2$ we have that $L_\gamma$ exceeds $L_*$ and disorder starts to control the behaviour, leading to a finite diffusion constant at $\gamma=0$. This can be seen by comparing Fig.~\ref{fig3}a, where for small dephasing $D$ reaches an upper limit given by the $\gamma=0$ case, whereas in the clean system in Fig.~\ref{fig2} diffusion constant can get arbitrarily large for a sufficiently small $\gamma$. Interestingly, we have $\lim_{\gamma \rightarrow 0}\lim_{L\rightarrow \infty} D(\gamma,L) \sim 1/h^{0.66}$, whereas on the other hand $\lim_{h \rightarrow 0}\lim_{L\rightarrow \infty} D(\gamma,L) \sim 1/\gamma^{0.33}$ -- the limits $\gamma \to 0$ and $h \to 0$ do not commute and depending on whether $\gamma$ is smaller or larger than $h^2$ diffusion can be either disorder or dephasing-dominated, respectively. Regardless of that, $D(\gamma)$ is always a decreasing function of dephasing strength $\gamma$.

\subsection{Case (ii) $h > h_{c2}$}
\begin{figure} 
  \includegraphics*[width=0.95\columnwidth]{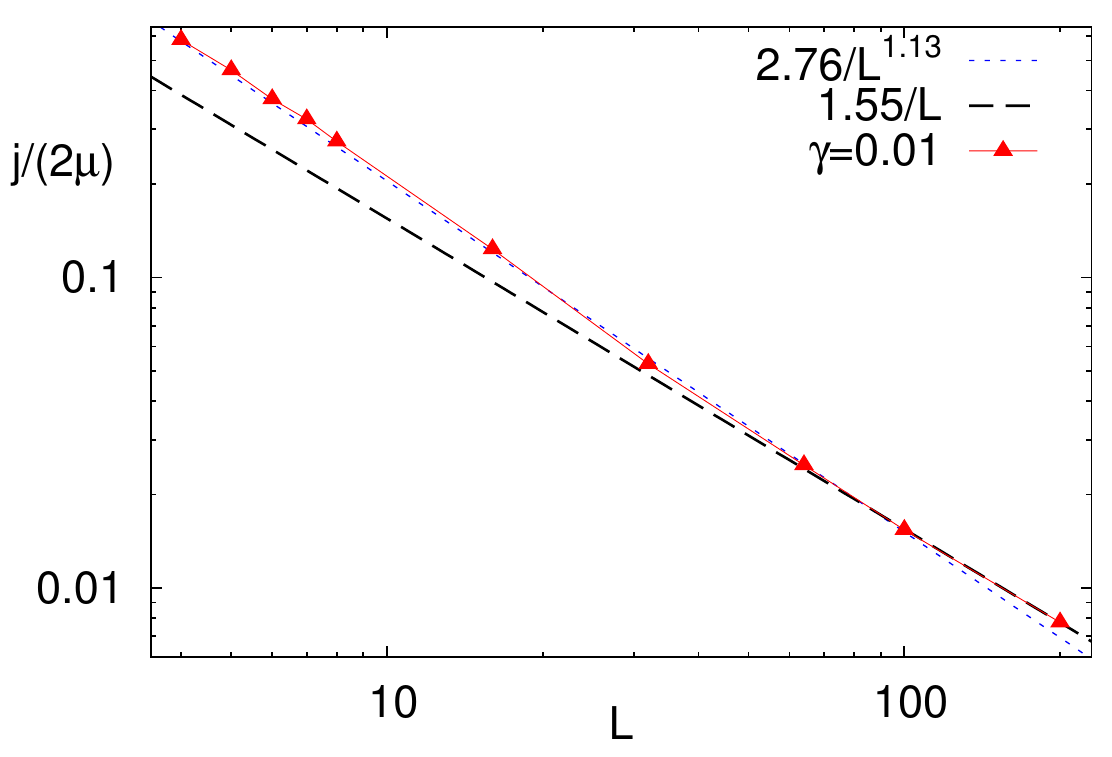} 
  \caption{The current $j$ is shown as a function of $L$ for $h=1.5$ (i.e., $h > h_{\rm c2}$) and $\gamma=0.01$. Compared to the $h<h_{\rm c2}$ case (Fig.~\ref{fig3}), now for $L<L_\gamma\approx 60$ transport is sub-diffusive ($\nu\approx 1.13$ in the figure).}
  \label{fig4}
\end{figure}
For strong dephasing there is not much difference compared to the case $h<h_{\rm c2}$, since $D(\gamma) \sim 1/\gamma$. At small dephasing though, when $L_\gamma$ is large, the system will for $L \ll L_\gamma$ behave as there would be no dephasing, i.e., in a sub-diffusive way, and will only asymptotically for $L \gg L_\gamma$ become diffusive as is demonstrated in Fig.~\ref{fig4}. The model without dephasing is sub-diffusive in this regime, i.e., $\nu>1$, for small $\gamma$ the diffusion constant as given by Eq.~\eqref{eq:smallg} will vanish when $\gamma \to 0$. Here one therefore has $\lim_{\gamma \rightarrow 0}\lim_{L\rightarrow \infty} D(\gamma,L) = 0$, compatible with sub-diffusion for $h>h_{\rm c2}$ and $\gamma=0$. Due to this a non-monotonic behaviour of $D(\gamma)$ is found, which must be contrasted with $D(\gamma)$ for $h<h_{\rm c2}$ that is always a monotonic function of $\gamma$.  

\begin{figure} 
  \includegraphics*[width=0.95\columnwidth]{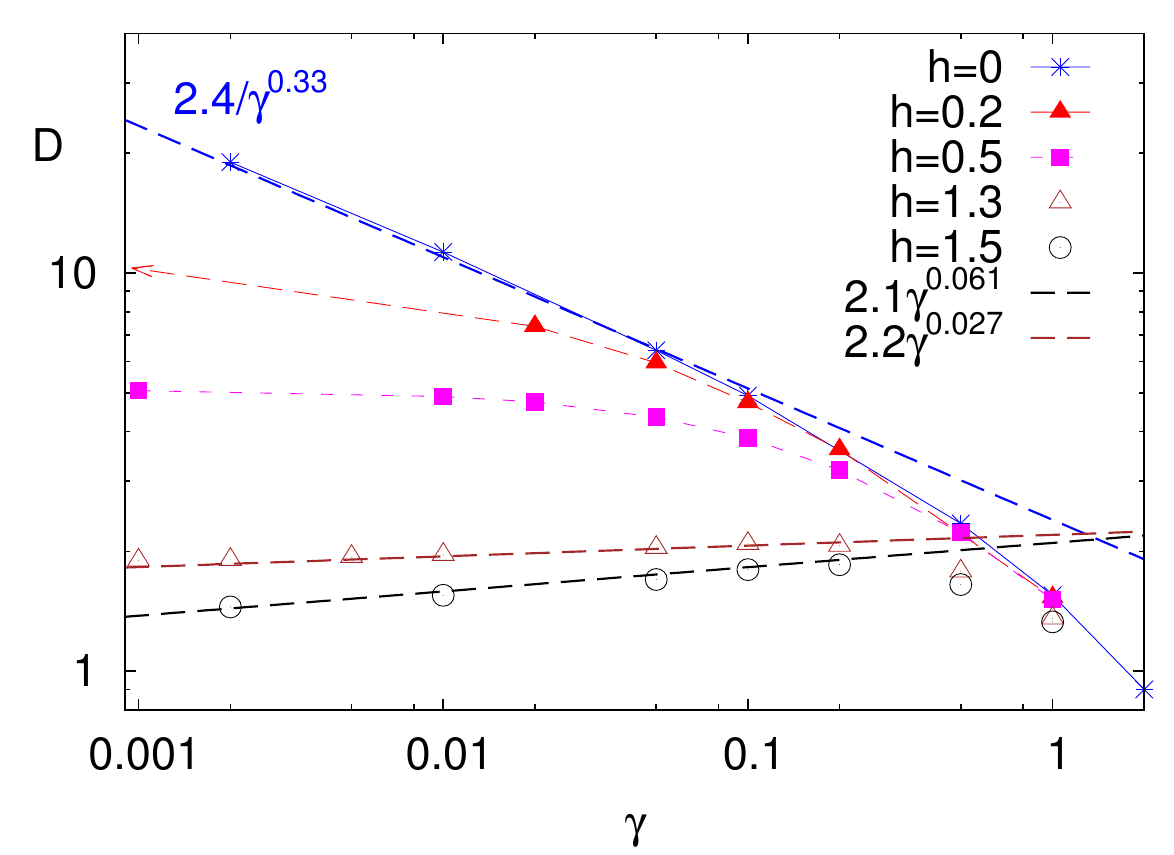} 
  \caption{Log-log plot of the dependence of the diffusion constant $D(\gamma)$ found for the largest system size $L$. Without disorder (blue stars) $D(\gamma \to 0)\to \infty$, with disorder and $h<h_{\rm c2}$ (diffusive noisless dynamics) $D(\gamma)$ increases for decreasing $\gamma$ (full symbols), saturating however at a finite $D(\gamma=0)$. For $h>h_{\rm c2}$ (sub-diffusive noisless dynamics) on the other hand $D(\gamma)$ decreases (empty symbols), leading to $D(\gamma \to 0) \to 0$ (two dashed lines are Eq.(\ref{eq:smallg}) using $\nu$ of the noiseless system read from e.g. small-$L$ slope in Fig.~\ref{fig4}).}
  \label{fig5}
\end{figure}

This is illustrated in Fig.~\ref{fig5} where we plot $D(\gamma)$ for different values of disorder $h$. We can see that, even though the spin transport is always diffusive for any $\gamma > 0$, a remnant of the diffusive to sub-diffusive transition at $h_{c2}$ found for $\gamma = 0$ is found to be embedded in the dependence of $D(\gamma)$. For $h<h_{c2}$ we find that $\lim_{\gamma \rightarrow 0}\lim_{L\rightarrow \infty} D(\gamma,L)$ is finite, while for $h>h_{c2}$ it is zero. This constitutes the main result of this work. The dependence $D(\gamma,h)$ is schematically illustrated in Fig.~\ref{fig:shema}. Therefore, in an experimental situation, where small dephasing might always be present, one can infer the transport properties of a model without dephasing by observing how $D$ varies as $\gamma$ decreases. If it grows one deals with a super-diffusive transport, if it saturates we have diffusion, and if it decreases one has sub-diffusion. For $h>h_{\rm c2}$ the diffusion constant has a maximum at an intermediate dephasing strength -- adding dephasing to a system whose transport is slower than diffusive, e.g., sub-diffusive, can increase transport.
  
\begin{figure} 
  \includegraphics*[width=0.95\columnwidth]{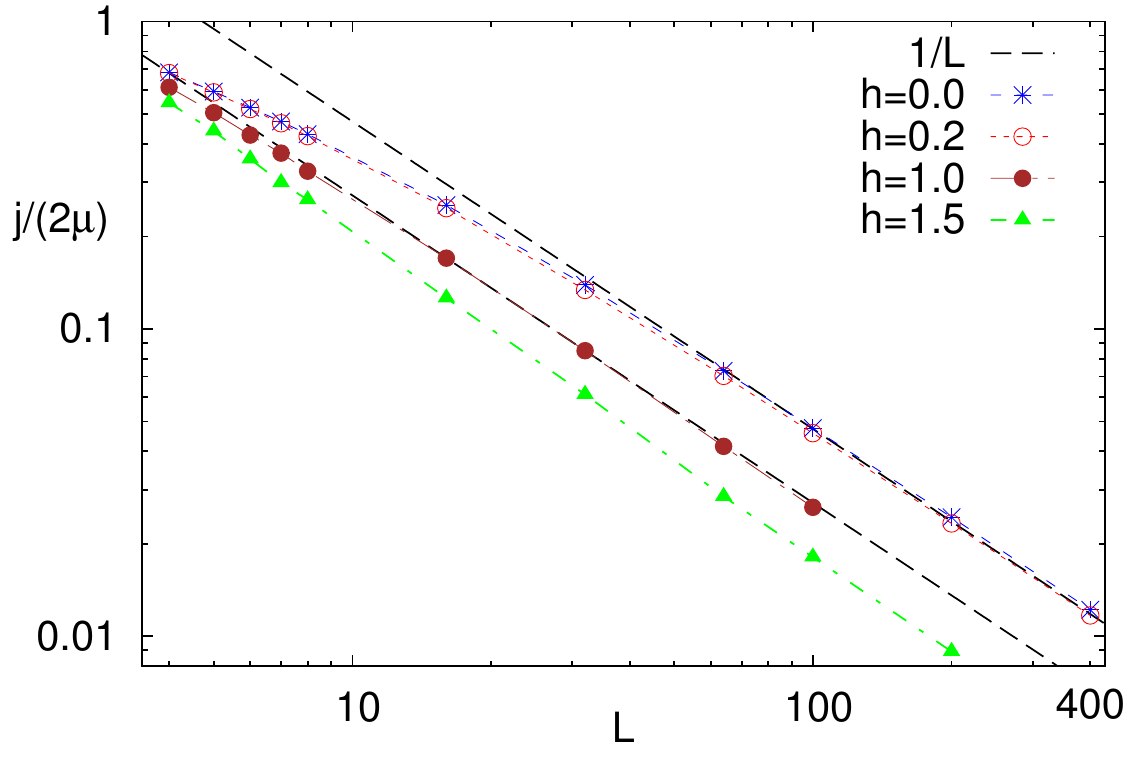} 
  \caption{Dependence of current on $L$ for different disorder strengths $h$ at a fixed dephasing $\gamma=0.1$. Current always decreases by increasing disorder.}
  \label{fig4b}
\end{figure}

\begin{figure} 
  \centerline{\includegraphics*[width=0.85\columnwidth]{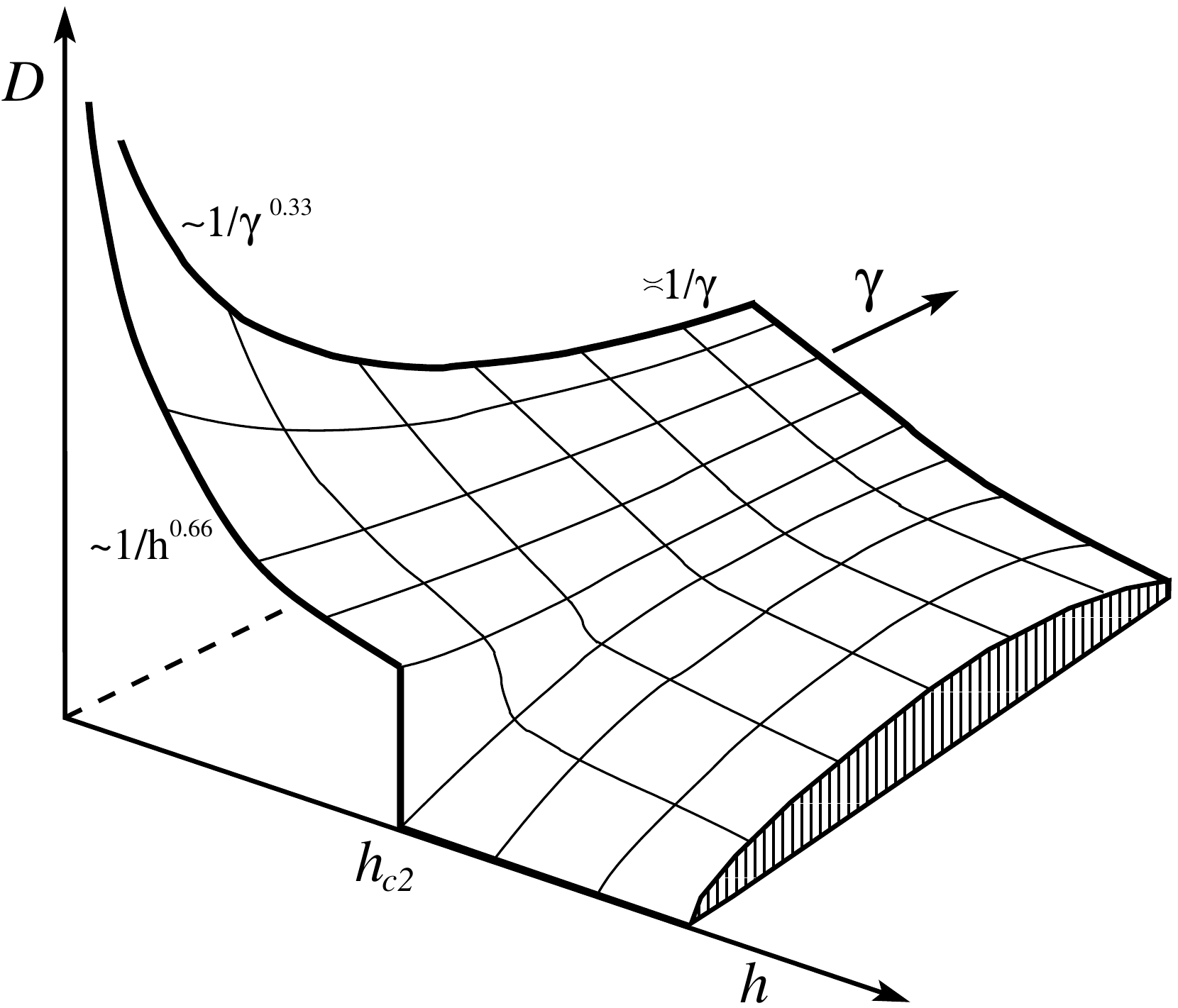}}
  \caption{Schematic diagram of the dependence of diffusion constant $D$ on dephasing strength $\gamma$ and disorder strength $h$. As described in Section~\ref{clean}, for the clean system there are two different asymptotic tendencies of $D$ with $\gamma$, namely $\sim\gamma^{-0.33}$ for low dephasing and $\sim\gamma^{-1}$ for large dephasing.}
  \label{fig:shema}
\end{figure}

One can also ask whether adding disorder to an otherwise clean system could help in increasing transport? For the model studied this can not happen because disorder, if it is sufficiently strong, always ``pushes'' transport towards localization, i.e., makes transport slower. This is confirmed by data in Fig.~\ref{fig4b} where we show $j(h,L)$ for fixed $\gamma=0.1$, where one observes that all current profiles lie below the $h=0$ curve.

\section{Conclusions}

Physical systems are never perfectly isolated from the environment. Therefore, an important question is how (small) coupling to external degrees of freedom modifies dynamics. In the present work we have focused on an ergodic phase of the isotropic one-dimensional Heisenberg model with disorder. In particular we studied spin transport in the presence of Markovian dephasing noise. We first identified relevant length scales in a system without disorder, which helped us to theoretically explain behaviour in the presence of both disorder and dephasing. We find that even though asymptotically dephasing noise always induces spin diffusion, this can happen in a nontrivial way. For small dephasing there is a remnant of the noiseless transport behaviour in the way that the diffusion constant varies with dephasing strength (Fig.~\ref{fig:shema}). An obvious extension of our work would be the generalisation to the anisotropic model with $\Delta<1$, or $\Delta>1$, where we expect similar behaviour as here, with possible interesting features for $h \to 0$. 

\section*{Acknowledgements} 
M.\v{Z} acknowledges Grant No. J1-7279 from the Slovenian
Research Agency. J.G. would like to thank V.~Varma and A.~Scardicchio for numerous discussions. The authors would like to acknowledge the use of the University of Oxford Advanced Research Computing (ARC) facility in carrying out this work. http://dx.doi.org/10.5281/zenodo.22558. This research is partially funded by the European Research Council under the European Union's Seventh Framework Programme (FP7/2007-2013)/ERC Grant Agreement no. 319286 Q-MAC. This work was also supported by the EPSRC National Quantum Technology Hub in Networked Quantum Information Processing (NQIT) EP/M013243/1. J.J.M.A. acknowledges financial support from Facultad de Ciencias at UniAndes-2015 project "Quantum control of non-equilibrium hybrid systems-Part II".

\bibliographystyle{andp2012}
\bibliography{dephase}

\end{document}